\documentclass{PoS}

\newcommand{\beqn}{\begin{eqnarray}}
\newcommand{\eeqn}{\end{eqnarray}}
\newcommand{\eq}[1]{(\ref{#1})}
\newcommand{\dd}{{\mathrm{d}}}
\newcommand{\dD}{{\mathrm{D}}}

\newcommand{\cZ}{{\mathcal Z}}

\newcommand{\Z}{{\mathbb{Z}}}
\newcommand{\Tr}{{\mathrm{Tr}}\,}

\title{
\vspace*{-1cm}
\begin{minipage}{\textwidth}
\begin{flushright}
\texttt{\footnotesize
PoS(LATTICE 2007)174\\%
ITEP-LAT/2007-05\\%
}
\end{flushright}
\end{minipage}\\[15pt]
Topological defects and equation of state \newline of gluon plasma}

\ShortTitle{Topological defects and equation of state of gluon plasma}

\author{\speaker{M. N. Chernodub}\\
ITEP, B. Cheremushkinskaya 25, Moscow, 117218, Russia}

\author{Katsuya Ishiguro\\
Institute for Theoretical Physics,
Kanazawa University, Kanazawa 920-1192, Japan
\\
RIKEN, Radiation Laboratory, Wako 351-0158, Japan}

\author{Atsushi Nakamura
\\
RIISE, Hiroshima University, Higashi-Hiroshima, 739-8527, Japan}

\author{Toru Sekido
\\
Institute for Theoretical Physics,
Kanazawa University, Kanazawa, 920-1192, Japan
\\
RIKEN, Radiation Laboratory, Wako 351-0158, Japan}

\author{Tsuneo Suzuki
\\
Institute for Theoretical Physics,
Kanazawa University, Kanazawa, 920-1192, Japan
\\
RIKEN, Radiation Laboratory, Wako 351-0158, Japan}

\author{V. I. Zakharov
\\
INFN, Sezione di Pisa, Largo Pontecorvo 3, 56127, Pisa, Italy
\\
ITEP, B. Cheremushkinskaya 25, Moscow, 117218, Russia}

\abstract{We show that the degrees of freedom associated with
magnetic monopole-- and vortexlike gluonic configurations make
a strong contribution to the anomaly of the energy--momentum tensor
of Yang--Mills theory in the deconfinement phase immediately above
the critical temperature. As is well known in zero--temperature
Yang--Mills theory, the monopoles and vortices are constituents of
a generic gluonic object in which the two neighbor monopoles are
connected together by a segment of vortex string. Our results provide
evidence that the monopole--vortex chains in SU(2) gauge theory and
their SU(3) counterparts, the monopole--vortex nets, are
thermodynamically relevant degrees of freedom in the gluonic plasma.}

\FullConference{The XXV International Symposium on Lattice Field Theory\\
         July 30-4 August 2007\\
         Regensburg, Germany}

\begin{document}

\section{Introduction}

The properties of thermal quark--gluon plasma in QCD have attracted great interest in recent
years~\cite{plasmareview,latticereview}. The plasma can be studied by both heavy--ion collision
experiments and numerical lattice simulations. The conventional theoretical approach to thermal
plasma is to treat it, in a zero approximation,
as a gas of free gluons and quarks supplemented with perturbative corrections. On the theoretical side,
the bulk characteristics of the plasma such as pressure and energy density
can then be represented in terms of perturbative series in the effective coupling
constant $g^{2}(T)$. The perturbative predictions for bulk quantities
turn to be in reasonable agreement with the available lattice data~\cite{latticereview}
for sufficiently high temperatures.

On the other hand, some particular properties of the plasma such as
viscosity~\cite{viscosity} indicate that in the zero approximation
the plasma at temperatures slightly above the critical temperature $T_c$
can be considered as an ideal liquid rather than an ideal gas.
There is not yet a coherent picture that unifies both perturbative
and nonperturbative features of the QCD plasma.

It was speculated in Refs. \cite{ref:PRL,chris} that there is a magnetic
component of Yang-Mills plasma that is crucial for determining the
plasma properties. In Ref.~\cite{chris} the constituents of the magnetic component
are thought to be classical magnetic monopoles. In Ref. \cite{ref:PRL}
the magnetic component is identified with so-called magnetic strings
which join (nonclassical) monopoles constituting chainlike
structures. The Abelian monopoles and the
center vortices are constituents of a generic gluonic object in which the two neighbor
monopoles are connected together by a segment of the vortex~\cite{chains,greensite}.
In $SU(2)$ gauge theory this object is considered as a monopole-vortex chain, while in the
$SU(3)$ case the objects form the monopole-vortex 3-nets.
The formation of the chains and nets is essential for the self--consistent
treatment of the monopoles in the quark-gluon plasma~\cite{ref:PRL}.

Both the magnetic (center) strings
and the (Abelian) monopoles as well as their role in the color confinement have been
discussed in the lattice community for more than a decade~\cite{greensite,review:monopoles}.
The properties of these defects change markedly once the temperature is increased above
the critical value $T_c$. In particular, these defects become predominantly time--oriented in accordance with the
assumption that they become a light component of the thermal gluon plasma~\cite{ref:PRL}.

Once the magnetic component of the plasma is identified with the topological defects,
further information on its properties can be obtained by direct numerical calculations on the
lattice at a finite temperature. Below we report the results of the first lattice measurements of the
contribution of the magnetic strings and magnetic monopoles to the equation of state
of the thermal Yang-Mills plasma.

\section{Equation of state and trace anomaly}

The free energy $F$ of the gauge system is expressed via a partition function $\cZ$ as follows:
\beqn
F = - T \log \cZ(T,V)\,,
\qquad
\cZ = \int \dD A \, \exp\left\{- \frac{1}{2 g^2} \Tr \, G_{\mu\nu}^2 \right\}\,,
\eeqn
where $G_{\mu\nu} = G_{\mu\nu}^a t^a$ is the field strength tensor of the non-Abelian field $A$ and
$t^a$ are the generators normalized in the standard way, $\Tr t^a t^b = \frac{1}{2} \delta^{ab}$.
The pressure $p$ and the energy density $\epsilon$ are given by the derivatives of the partition function
with respect to the spatial volume of the system and with respect to the temperature:
\beqn
p = \frac{T}{V} \frac{\partial \log Z(T,V)}{\partial \log V} = - \frac{F}{V} = \frac{T}{V} \log \cZ(T,V)\,, \qquad
\varepsilon = \frac{T}{V} \frac{\partial \log Z(T,V)}{\partial \log T}\,.
\label{eq:pressure}
\eeqn
The last two equalities for the pressure are valid for a sufficiently large and homogeneous system in
thermodynamical equilibrium. The relation between the pressure and the energy in Eq.~\eq{eq:pressure} constitutes
the equation of state of the system.

According to Eq.~\eq{eq:pressure} it is sufficient to determine the partition function of the system to calculate the
corresponding equation of state. However, lattice simulations are suitable for calculation of quantum averages of
operators rather than the partition function itself. On the other hand, both the energy and the pressure can be derived
from the quantum average of a single quantity, which is the trace of the energy--momentum tensor $T_{\mu\nu}$.

In $SU(N)$ gauge theory the energy--momentum tensor is given by the formula
\beqn
T_{\mu\nu} = 2 \Tr \left[G_{\mu\sigma} G_{\nu\sigma} - \frac{1}{4} \delta_{\mu\nu} G_{\sigma\rho} G_{\sigma\rho}\right]\,,
\label{eq:T}
\eeqn
which is traceless because the {\it bare} Yang--Mills theory is a conformal theory. However, because of a dimensional
transmutation the conformal invariance is broken at the quantum level and the energy--momentum tensor exhibits
a trace anomaly. The thermodynamic relations in Eq.~\eq{eq:pressure} give~us
\beqn
\theta(T) = \langle T^\mu_\mu \rangle \equiv \varepsilon - 3 p = T^5 \frac{\partial}{\partial T} \frac{p(T)}{T^4}
= - T^5 \frac{\partial}{\partial T} \frac{\log \cZ(T,V)}{T^3 V}\,.
\label{eq:anomaly:continuum}
\eeqn
The pressure and energy density can be expressed via the trace anomaly as follows:
\beqn
p(T) = T^4 \int\limits^T \ \frac{\dd\, T_1}{T_1} \ \frac{\theta(T_1)}{T_1^4}\,, \qquad
\varepsilon(T) = 3 \, T^4 \int\limits^T \ \frac{\dd\, T_1}{T_1} \ \frac{\theta(T_1)}{T_1^4} + \theta(T)\,.
\label{eq:pressure:anomaly}
\eeqn
Thus the trace anomaly is a key quantity that allows us to reconstruct the whole equation of state.

Note that the trace anomaly should vanish in the case of free relativistic particles
($\varepsilon = 3 p$), or in the case when excitations are
too massive compared with the temperature, $m \gg T$
(then $\varepsilon \sim p \sim \exp\{- m/T\}$). For Yang--Mills theory
these statements imply that the dimensionless quantity $\theta/T^4$ should approach zero at both
high temperatures (the gluons form a weakly interacting gas) and low temperatures
(the mass gap is much greater than the temperature).

%\section{Trace anomaly of pure gluons}

The partition function of $SU(N)$ lattice gauge theory is written in the Wilson form,
\beqn
\cZ(T,V) = \int D U \, \exp\Bigl\{ - \beta \sum_P S_P[U]\Bigr\}\,, \qquad S_P[U] = 1 - \frac{1}{N} {\mathrm{Re}}\, \Tr U_P\,.
\label{eq:lattice:cZ}
\eeqn
The temperature $T = 1/(N_t a)$ and the volume $V = (N_s a)^3$ of the system are related to the geometry of
the asymmetric lattice, $N_s^3 N_t$ and to the lattice spacing $a$ which is a function of the lattice
coupling $\beta = 2 N /g^2$. Using the relation $T (\partial/\partial T) = - a (\partial/\partial a)$ one can
adopt Eq.~\eq{eq:anomaly:continuum} to describe the lattice thermodynamics,
\beqn
\frac{\theta(T)}{T^4} = 6 \, N_t^4 \left(\frac{\partial \beta(a)}{\partial \log a} \right)
\cdot \left(\langle S_P \rangle_T - \langle S_P \rangle_0\right)\,,
\label{eq:anomaly:lattice}
\eeqn
where the plaquette averages ${\langle S_P \rangle}_T$ and ${\langle S_P \rangle}_0$ are taken,
respectively, in the thermal bath at $T>0$ and in the zero--temperature case corresponding to
the asymmetric $N_s^3 N_t$ and symmetric $N_s^4$ lattices. In Eq.~\eq{eq:anomaly:lattice}
it is implied that the $T=0$ plaquette expectation value is subtracted to remove the effect
of quantum fluctuations, which lead to ultraviolet divergency of the quantum expectation
value. As a result, the trace anomaly becomes an ultraviolet quantity, which is normalized
to zero at $T=0$ because of the existence of the mass gap. The trace anomalies and
the equation of states for $SU(2)$ and $SU(3)$ gauge theories were calculated in Refs.~\cite{Engels:1988ph} and \cite{Boyd:1996bx},
respectively.

\section{Trace anomaly from monopoles in SU(3) lattice gauge theory}

The magnetic monopoles are particle--like configurations appearing as singularities in the diagonal
component of the gluonic field $A^{\mathrm{diag}}_\mu$ in the so--called Maximal Abelian gauge, which makes the off-diagonal
gluon components $A^{\mathrm{off}}_\mu$ non--propagating~\cite{ref:tHooft} (a review can be found in Ref.~\cite{review:monopoles}).
The local gauge condition can formally be
written as $D^{\mathrm{diag}}_\mu A^{\mathrm{off}}_\mu = 0$. The monopole trajectories $k_\mu$
are identified as sources of the Abelian magnetic field in the diagonal component of the gauge field,
$k_\mu \sim \partial_\nu {\tilde F}^{\mathrm{diag}}_{\mu\nu}$. The magnetic charge is quantized and
conserved quantity. Details of gauge fixing and the determination of the monopole trajectories
on the lattice can be found in Ref.~\cite{ref:details:SU3}.

The partition function of Yang--Mills theory can be represented as a product of two parts: the first
part is from the monopole contribution, while the second part is given by
the remaining field fluctuations including a perturbative contribution. The
trace of the energy--momentum tensor in Eq.~\eq{eq:anomaly:continuum} can consequently be represented
as the sum of these parts,
\beqn
\cZ = \cZ^{\mathrm{mon}} \cZ^{\mathrm{rest}}\,,\qquad \theta = \theta^{\mathrm{mon}}(T) + \theta^{\mathrm{rest}}(T)\,.
\label{eq:separation}
\eeqn

\begin{figure}[!htb]
\begin{center}
\includegraphics[angle=-0,scale=0.8,clip=true]{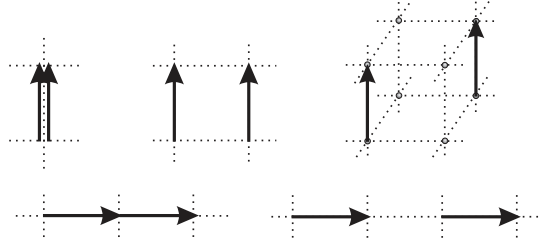}
\end{center}
\vskip -5mm
\caption{Schematic representation of typical terms in the lattice monopole action.}
\label{fig:monopole:interactions}
\end{figure}

The monopole partition function is given by the sum over the closed monopole trajectories,
\beqn
\cZ^{\mathrm{mon}} = \sum_{\delta k = 0} \exp\left\{ - \sum_i f_i(\beta) S^{\mathrm{{mon}}}_i(k)\right\}\,,
\qquad
S^{\mathrm{{mon}}}_i(k) = \sum_{s,\mu} \sum_{s',\nu} k_{\mu}(s) K^{(i)}_{\mu\nu}(s,s') k_{\nu}(s')\,,
\label{eq:Zmon}
\eeqn
where the monopole action consists of the two-point interaction terms $S^{\mathrm{{mon}}}_i$ between
the elementary segments of the closed ($\delta k = 0$) monopole trajectories.
Some part of the interactions terms -- defined by the kernels $K^{(i)}_{\mu\nu}(s,s')$ in Eq.~\eq{eq:Zmon} --
are shown in Figure~\ref{fig:monopole:interactions}.

The coupling constants of the monopole action $f_i$ are numerically determined as functions of $\beta$ with
the help of the inverse Monte Carlo (MC) method~\cite{ref:details:SU3,ref:ShibaSuzuki}. The inverse MC algorithm
uses an
ensemble of the monopole trajectories as the input, which are located in the original gluonic
configurations by the Abelian projection method. We used 400 statistically independent
configurations of the $SU(3)$ gauge field generated by the
usual MC procedure for $L_s=16$ and $L_t=4,16$ lattices.
The inverse MC algorithm searches for the best monopole action that can most consistently
describe the ensemble of the monopole trajectories by the partition function in Eq.~\eq{eq:Zmon}.
In our simulations we truncated the monopole action after 11 terms allowed us to describe the
available ensembles of the monopole trajectories with acceptable accuracy.

Having determined the monopole action we can calculate the contribution of the magnetic monopoles to the
trace anomaly of the gluon plasma,
\beqn
\theta^{\mathrm{mon}} = N^4_t \left(a \frac{\partial \beta}{\partial a}\right) \sum_i
\left(\frac{\partial f_i(\beta)}{\partial \beta}\right)
\left[\langle {\bar S}^{\mathrm{mon}}_i\rangle_T - \langle {\bar S}^{\mathrm{mon}}_i\rangle_0\right]\,,
\qquad
{\bar S}^{\mathrm{mon}}_i = \frac{1}{N_s^3 N_t} S^{\mathrm{mon}}_i\,,
\label{eq:anomaly:monopoles}
\eeqn
in which we used Eqs.~\eq{eq:anomaly:continuum}, \eq{eq:separation} and \eq{eq:Zmon}. Analogously
to Eq.~\eq{eq:anomaly:lattice} we have normalized the finite--temperature expectation values
of the parts $S^{\mathrm{mon}}_i$ of the monopole action, shifting these expectation values by the
corresponding zero--temperature expectation values. The subtraction procedure is required
to remove the ultraviolet divergencies and to correctly normalize the monopole trace anomaly.

\begin{figure}[!htb]
\begin{center}
\begin{tabular}{cc}
\includegraphics[angle=-0,scale=0.32,clip=true]{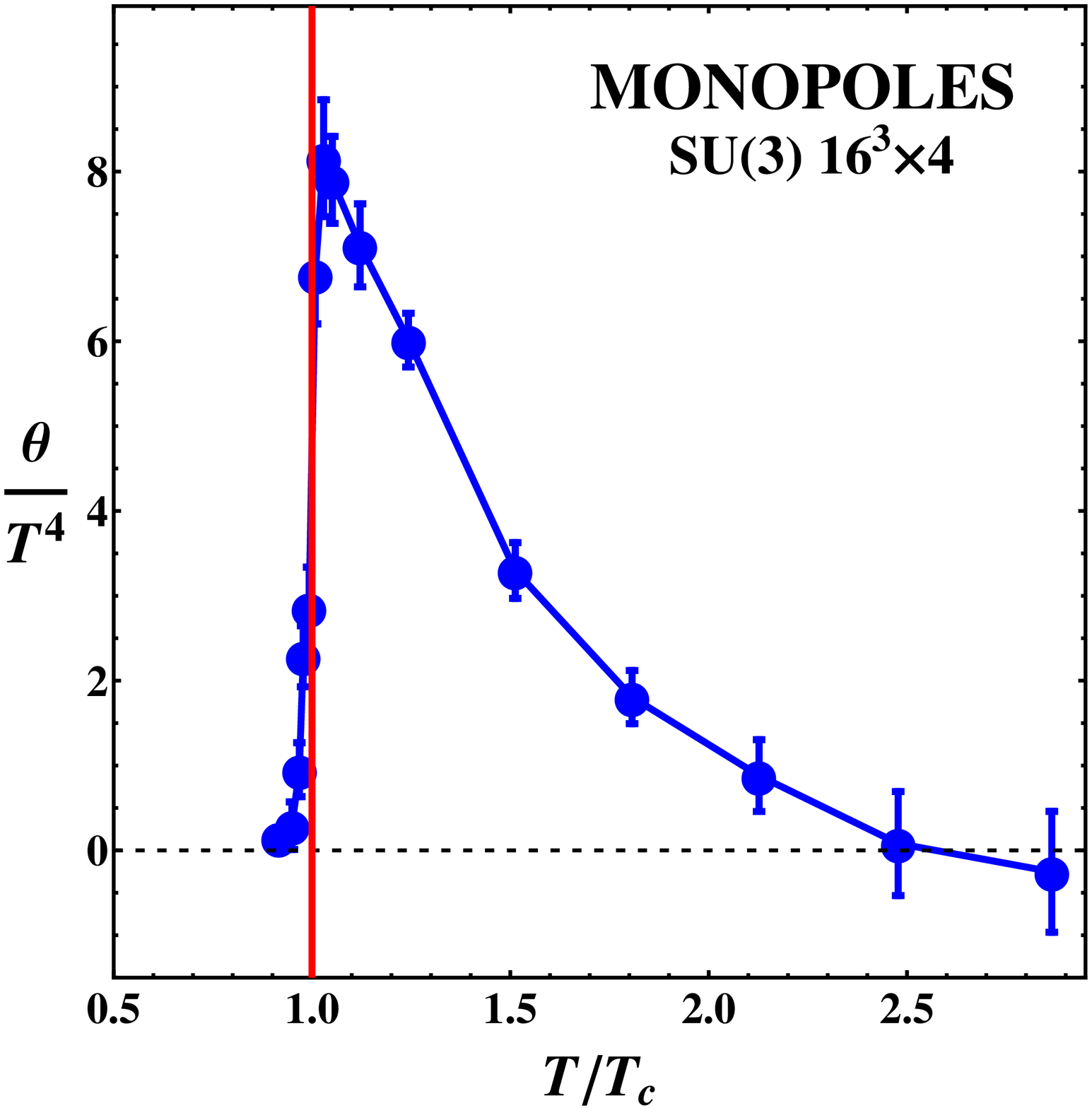}
&
\includegraphics[angle=-0,scale=0.327,clip=true]{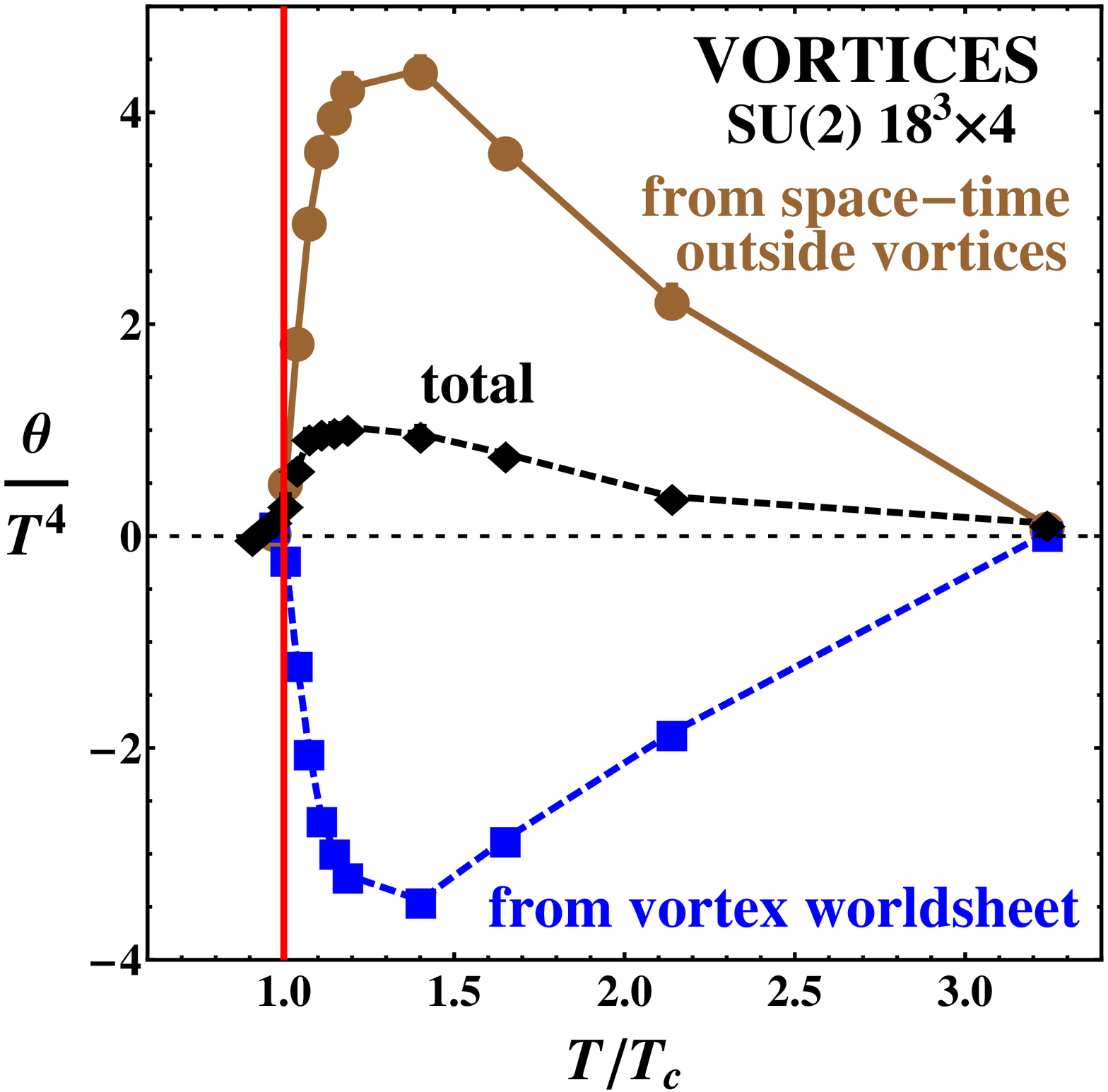} \\
\hskip 7mm (a) & \hskip 8mm (b)
\end{tabular}
\end{center}
\vskip -5mm
\caption{(a) Contribution of the magnetic monopoles to the trace anomaly in $SU(3)$ lattice gauge theory.
(b) The trace anomaly in $SU(2)$ lattice gauge theory is divided into contributions from the vortex worldsheets
(squares) and from the whole space-time located outside the center vortices (circles). Their sum gives the total contribution
(diamonds). The lines connecting the data points are drawn to guide the eye. The vertical lines mark the deconfinement
temperature $T_c$.}
\label{fig:anomaly}
\end{figure}

The contribution of the monopoles to the trace anomaly $\theta/T^4$ in $SU(3)$ lattice gauge
theory is shown in Figure~\ref{fig:anomaly}(a). In the confinement region the trace anomaly
is zero. The anomaly starts to increase at $T \sim T_c$ approaching a maximum at a temperature
slightly above the deconfinement temperature\footnote{Note that this maximum is higher than that for the pure gluons
[Eq.~\eq{eq:anomaly:lattice}] as calculated in Ref.~\cite{Boyd:1996bx}. We attribute this difference to the large
finite--size corrections originating from the relatively small and rough lattice ($N_t=4$) used in our numerical calculations.
To improve our results at the quantitative level, one should check the scaling towards the continuum limit
(this work is currently in preparation).}. One also notices that the
monopole contribution to the anomaly is a positive quantity that
increases at slower rate than $T^4$ for $T \gg T_c$. All these properties qualitatively match
those of the original gluonic trace anomaly in Eq.~\eq{eq:anomaly:lattice}
calculated numerically in Ref.~\cite{Boyd:1996bx}. We conclude that the monopoles do contribute to the equation of state of the gluon plasma.

\section{Trace anomaly from vortex worldsheets in SU(2) gauge theory}

The vortices are magnetic defects, which can be identified in gluon field configurations using the
so--called Maximal Center gauge (a review and details can be found in Ref.~\cite{greensite}).
The gauge fixes the non-Abelian gauge group up to its center subgroup. The magnetic vortices are
then considered as stringlike defects in the center gauge variables.

In $SU(2)$ lattice gauge theory the position of the vortex is determined using
the $\Z_2$ gauge field $Z_l = {\mathrm{sign}} \Tr U_l = \pm 1$.
The lattice field--strength tensor of the $\Z_2$
gauge field, $Z_P = \prod_{l \in \partial P} Z_l$, takes the negative
value $Z_P = -1$ if the plaquette $P$ is pierced by the vortex worldsheets ${}^*\sigma_{\mu\nu}(s)$
on the dual lattice. If $Z_P = +1$ then the  plaquette $P$ is not pierced
by the vortex.

The gluonic trace anomaly in Eq.~\eq{eq:anomaly:lattice} is proportional to
the expectation values of the plaquette action. Therefore, the separation of
space-time into two subspaces (occupied and not occupied by vortices) leads to the
natural splitting of the gluonic contribution to the trace anomaly into that
originating from the vortex worldsheets,
$\theta_{\mathrm{vort}}$, and that from elsewhere, $\theta_{\mathrm{rest}}$:
\beqn
\theta = \theta_{\mathrm{vort}} + \theta_{\mathrm{rest}} \quad \mbox{since} \quad
\langle S_P \rangle = \langle S_P \rangle_{\mathrm{vort}} + \langle S_P \rangle_{\mathrm{rest}}
\quad \mbox{and} \quad
\sum_P S_P =  \sum_{P \in \sigma} S_P + \sum_{P \not\in \sigma} S_P\,.
\eeqn
The two contributions to the plaquette action can be conveniently written as
\beqn
\langle S_P \rangle_{\mathrm{vort}} = \frac{1}{N_P} \langle \sum_{P \in \sigma} S_P \rangle
= \frac{1}{2} \left(\langle S_P \rangle - \langle {\widetilde{S}}_P \rangle\right)\,, \qquad
\langle S_P \rangle_{\mathrm{rest}} = \frac{1}{N_P} \langle \sum_{P \not\in \sigma} S_P \rangle
= \frac{1}{2} \left(\langle S_P \rangle + \langle {\widetilde{S}}_P \rangle\right),
\eeqn
where $N_P = 6 N_s^3 \times N_t$ is the total number of plaquettes on the lattice. The action
\beqn
{\widetilde{S}}_P[U] = S_P[\widetilde{U}] = 1 - \frac{1}{2} \Tr \widetilde{U}_P =  1 - \frac{1}{2} Z_P \Tr U_P\,,
\quad \mbox{where} \quad \widetilde{U}_l = Z_l U_l\,, \quad Z_l = {\mathrm{sign}} \Tr U_l\,,
\eeqn
can be interpreted as the action of the system with formally ``removed'' vortices.
The above relations are valid in the Maximal Center gauge. The standard plaquette
action $S_P[U]$ is given by Eq.~\eq{eq:lattice:cZ}.

In Figure~\ref{fig:anomaly}(b) we show the both contributions to the trace anomaly calculated
for $L_s=18$ and $L_t=4,18$ lattices
using from 100 to 800 configurations (depending on the value of $\beta$ and the lattice
geometry). The contribution from the vortex worldsheets is negative, in agreement with
general theoretical expectations~\cite{Gorsky:2007bi}. The maximum absolute value of the vortex
contribution is about three times larger than the pure--gluon contribution
calculated numerically in Ref.~\cite{Engels:1988ph}. This property of the topological magnetic
contribution agrees with the observation~\cite{mitrjushkin} that the magnetic gluon condensate
provides a large negative contribution to the equation of state.
Finally, the contribution to the anomaly originating from the rest of the space-time
(outside the vortex worldsheets) is large and positive. The total sum
of these contributions [also shown in Figure~\ref{fig:anomaly}(b)]
is in agreement with a known result in the $SU(2)$ lattice gauge theory~\cite{Engels:1988ph}.

It is worth noting that for the chosen lattice parameters the maximal contribution of the magnetic
vortices to the trace anomaly is achieved when the vortices occupy on average only $5\%$ of the space-time.
The negative contribution from the vortices is almost canceled by the positive contribution from
the rest (95\%) of the space--time. This fact allows us to conclude that the local gluonic fields in the
vortex worldsheets are much stronger than the fields outside the vortices.

\section{Conclusion}

We found that both the magnetic defects, the monopoles and the vortices, contribute
significantly to the trace anomaly and, via Eq.~\eq{eq:pressure:anomaly}, to the
equation of state of the gluon plasma. The contribution of the monopoles is
positive while the vortices provide a negative contribution. These results are
particularly interesting in view of the fact~\cite{chains,greensite} -- which
is particularly important in the plasma regime~\cite{ref:PRL} --
that the monopoles and vortices are part of the generic object, which
constitutes a monopole--vortex chain/net.
The contribution from the monopoles is calculated through
the determination of the monopole action, which takes into account local self-interaction as well as
nonlocal interactions between separated monopoles. In contrast, the
contribution of the magnetic vortices to the anomaly is calculated locally. Thus, the qualitative
difference between the monopole and vortex contributions is most probably due to the effect of
the nonlocal interactions.

In conclusion, we stress our main result: the monopole-vortex chains in
$SU(2)$ gauge theory and the monopole--vortex nets in $SU(3)$ gauge theory are thermodynamically
relevant objects in gluon plasma, because they contribute significantly to the equation of state of the plasma.

This work was supported by Grants-in-Aid for Scientific Research from
``The Ministry of Education, Culture, Sports, Science and Technology''
Nos. 13135216 and 17340080, by grants RFBR 05-02-16306a and RFBR\--DFG 06-02-04010,
and by a STINT Institutional grant IG2004-2 025.
The numerical simulations were performed using
a SX7 supercomputer at RIKEN, SX5 and SX8 machines at RCNP
at Osaka University, and a SR11000 machine at Hiroshima
University.

\end{document}